\documentclass[letters,useAMS,usenatbib]{mnras}
\usepackage{psfig,url}
\usepackage{hyperref}
\usepackage[all]{hypcap}
\newcommand{\figdir}{.}
\newcommand{\mfigur}[3]
    {\begin{figure}
        \centerline{\psfig{figure=\figdir/#1.ps,height=#2,clip=}}
        \caption{\label{#1}#3}
    \end{figure}}
\newcommand{\be}[1]{\begin{equation}\label{#1}}
\newcommand{\ee}{\end{equation}}
\newcommand{\bea}[1]{\begin{eqnarray}\label{#1}}
\newcommand{\eea}{\end{eqnarray}}
\newcommand{\lwig}{{\leavevmode\kern0.3em\raise.3ex\hbox{$<$}
                    \kern-0.8em\lower.7ex \hbox{$\sim$}\kern0.3em}}
\newcommand{\dd} [2]{{{{\rm d}{#1}\over{\rm d}{#2}}}}

\newlength{\pwidth}
\newcommand{\ppmm}[2]{\lower-.7ex\hbox{\scriptsize$#1$}\settowidth{\pwidth}%
            {\scriptsize$#2$}\kern-\pwidth\lower.5ex\hbox{\scriptsize$#2$}}
\newcommand{\sfrac}[2]{{\leavevmode
                    \kern.1em\raise.5ex\hbox{\the\scriptfont0 #1}
                    \kern-.13em\raise.25ex\hbox{\the\scriptfont0 /}
                    \kern-0.12em\lower.25ex\hbox{\the\scriptfont0 #2}}}
\begin{document}
    \title[Surface effect of convective expansion of atmosphere]
          {The asteroseismic surface effect from a grid of\\
           3D convection simulations.  I. Frequency shifts from\\
           convective expansion of stellar atmospheres}
    \author[R. Trampedach, et al.]{Regner Trampedach$^{1,2}$\thanks{E-mail: rtrampedach@SpaceScience.org},
     Magnus J.\ Aarslev$^{2}$,
     G{\"u}nter Houdek$^{2}$,
     Remo Collet$^{2}$,     \newauthor
     J{\o}rgen {Christensen-Dalsgaard}$^{1}$,
     Robert F. Stein$^{3}$  and
     Martin Asplund$^{4}$\\
    $^{1}$Space Science Institute, 4750 Walnut Street, Suite 205,
          Boulder, CO 80301 USA\\
    $^{2}$Stellar Astrophysics Centre, Dept.\ of Physics and Astronomy,
          Ny Munkegade 120, Aarhus University, DK--8000 Aarhus C, Denmark\\
    $^{3}$Department of Physics and Astronomy, Michigan State University
          East Lansing, MI 48824, USA\\
    $^{4}$Research School of Astronomy and Astrophysics,
          Mt.\ Stromlo Observatory, Cotter Road, Weston ACT 2611, Australia}

    \date{Received \today / Accepted \today}

    \maketitle

    \begin{abstract}
We analyse the effect on adiabatic stellar oscillation frequencies of replacing
the near-surface layers in 1D stellar structure models with averaged 3D stellar
surface convection simulations. The main difference is an expansion of the
atmosphere by 3D convection, expected to explain a major part of the
asteroseismic surface effect; a systematic
overestimation of p-mode frequencies due to inadequate surface physics.

We employ pairs of 1D stellar envelope models and 3D simulations
from a previous calibration of the mixing-length parameter, $\alpha$.
That calibration constitutes the hitherto most consistent matching
of 1D models to 3D simulations, ensuring that their
differences are not spurious, but entirely due to the 3D nature of convection.
The resulting frequency shift is
identified as the \emph{structural} part of the surface effect.
The important,
typically non-adiabatic, \emph{modal} components of the surface effect are not
included in the present analysis, but relegated to future papers.

Evaluating the structural surface effect at the frequency of maximum
mode amplitude, $\nu_{\max}$, we find shifts from $\delta\nu=-0.8\,\mu$Hz for
giants at $\log g=2.2$ to $-35\,\mu$Hz for a
($T_{\rm eff}=6901$\,K, $\log g=4.29$) dwarf. The fractional effect
$\delta\nu(\nu_{\max})/\nu_{\max}$, ranges from $-0.1$\% for
a cool dwarf (4185\,K, 4.74) to $-6$\% for a warm giant (4962\,K, 2.20).
    \end{abstract}


\section{Introduction}
\label{sect:intro}

The asteroseismic surface effect is a systematic difference between measured
stellar p-mode frequencies and theoretical, adiabatic frequencies of stellar
models, known to arise from differences in the surface layers
\citep{brown:SurfEff,jcd-Nature:osc-eos}.

\citet{rosenthal:conv-osc} first analysed the helioseismic surface effect
in terms of frequency differences between 1D models and averaged 3D surface
simulations. They concluded that most of the effect is due to a convective
expansion of the atmosphere, compared to 1D convective atmospheres.
Similar analyses have now been carried out by \citet{piau:3D-SunSeism}
and \citet{magic:SolarSurfEff} for the
Sun, \citet{sonoi:Sun3DSurfEffect} for a grid of 10 simulations, and by
\citet{ball:SeismSurfExpanEff} for four simulations on the main sequence,
all for solar metallicity.

\citet{houdek:3DsunSurfEff} recently presented an analysis of various components
of the surface effect for the solar case. We use the same 3D solar simulation,
but extend our analysis to the whole grid of solar metallicity simulations
\citep{trampedach:3Datmgrid}, to explore the behaviour with atmospheric
parameters. On the other hand, we limit our analysis to only include the
stratification contributions to the seismic surface effect (see Section
\ref{sect:1D-3DSurfEff}), and defer the evaluation of modal components to
a future paper.

\section{The 3D convective atmosphere simulations}
\label{sect:3Dsims}

We use the grid of 37 fully compressible 3D radiation-coupled hydrodynamic
simulations of stellar surface convection by \citet{trampedach:3Datmgrid}. The grid covers effective temperatures,
$T_{\rm eff}=4$\,200--6\,900\,K on the main sequence, and $T_{\rm eff}=4$\,300--5\,000\,K and $\log g=2.2$--$2.4$
for giants, all for solar metallicity.
The simulations evolve the conservation
equations of mass, momentum and energy in a small box straddling the stellar
photosphere extending up to a logarithmic, Rosseland optical depth of $\log\tau=-4.5$ and reaching more than 6 pressure
scale heights below the photosphere. This is deep enough that they can be
safely merged with 1D models below, i.e., the convective fluctuations there are
small, as is the deviation from an adiabatic stratification.

The simulations employ a 15 element custom calculation of the so-called
Mihalas-Hummer-D{\"a}ppen equation of state (EOS) \citep{mhd1}, and
realistic monochromatic opacities \citep{trampedach:T-tau}.
An open bottom boundary mimics the effect of the large entropy reservoir of the
convection zone below the simulation, ensuring a realistic entropy
contrast inside the simulations.

\subsection{Consistently patched 1D models and 3D simulations}
\label{sect:patching}

The above simulations were previously used to calibrate 
the main parameter, $\alpha$, of the mixing-length
formulation (MLT) of 1D convection
\citep{trampedach:alfa-fit}.
This was carried out as a matching of temperature and density at the common total pressure of the
matching point. 
The matching point is chosen as deep in the
simulation as possible (less than a pressure scale-height from the bottom), while still avoiding boundary effects (See Fig.~\ref{match_check}).
The 1D models include a turbulent pressure (See Sect.~\ref{sect:Pturb}),
$p_{\rm t}^{\rm 1D}=\beta\varrho v_{\rm MLT}^2$, with
form-factor $\beta$. This pressure is
smoothly suppressed towards the surface, as the MLT version would give an
unphysically sharply peaked $p_{\rm t}^{\rm 1D}$. The approach is therefore to
include it at the fitting point and below, to ensure a consistent match to the
3D simulations, as shown in Fig.~\ref{match_check}, but suppress it before it
becomes important for the hydrostatic equilibrium
\citep[see][for details]{trampedach:alfa-fit}. Formulating a realistic
turbulent pressure for 1D models is a
separate project. We iterated for $\alpha$ and $\beta$ until both converged
to within $10^{-6}$ resulting in deviations of $\log T$ and $\log\varrho$ at
the matching point of less than $10^{-5}$ and $10^{-3}$, respectively.
The 1D models employ the exact same
EOS and abundances as the simulations, and in the atmosphere use the opacities
of the simulations, and the temperature stratification, $T(\tau)$, extracted from the simulations \citep{trampedach:T-tau}. This ensures that the 1D envelope model and the averaged
3D simulation can be patched together for a continuous and smooth model
across the matching point.
This is illustrated in Fig.~\ref{match_check} for the warmest dwarf simulation,
which deviates most strongly from its calibrated 1D model.
\mfigur{match_check}{7.2cm}
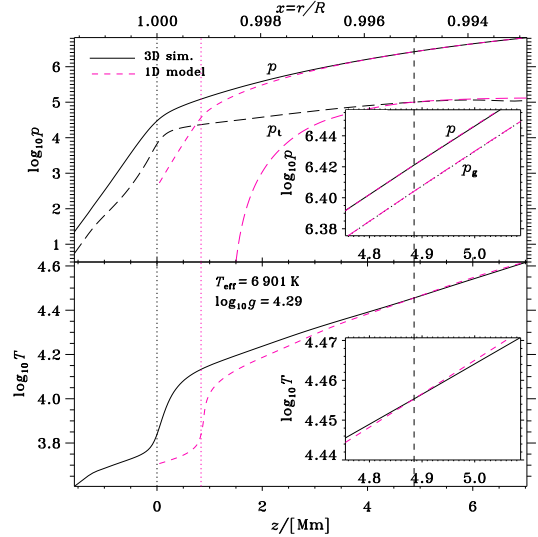
    {Top panel: the pressures (black) of the 3D simulation (also shown in
     Fig.~\ref{turb_expan}) and the
     calibrated 1D model (magenta). The insert panel is a zoom around the
     matching point (vertical, black dashed, shared with main panel).
     The surface (where $\langle T\rangle=T_{\rm eff}$) is shown for the
     3D simulation (vertical, black dotted) and the 1D model (magenta dotted),
     The difference causes the structural surface effect.
     Bottom panel: Same as top panel but for the logarithmic temperature.}

The 3D simulations are slightly more extended than the (un-patched) 1D models
(giving rise to the structural surface effect), and hence have slightly lower
$T_{\rm eff}$ and $\log g$, corresponding to their common mass and luminosity.
The simulations are carried out in the plane-parallel approximation (constant
surface gravity) and the averages are therefore corrected for sphericity
consistent with the radius of the 1D model, to avoid glitches at the
matching point. Of these steps, only the consistent EOS and
abundances have been implemented in previous work
\citep{piau:3D-SunSeism,sonoi:Sun3DSurfEffect,ball:SeismSurfExpanEff,magic:SolarSurfEff}.

The 1D models are computed with the stellar envelope
code by \citet{jcd-srf:stel-osc}, which is closely related to the
ASTEC stellar structure and evolution code \citep{jcd:ASTEC}. Being
envelope models, they ignore the innermost 5\% of the star, as well as nuclear reactions
and any other composition-altering processes. The limited extent constrains
the modes available for our analysis to those with turning points well
inside the envelope model. Our results will, however, not be affected, as the
surface effect is indeed confined close to the surface and fully contained
even in the 3D, deep atmosphere simulations.

\section{The structural part of the asteroseismic surface effect}
\label{sect:1D-3DSurfEff}

We compare adiabatic mode frequencies from two cases:

\noindent{\bf UPM:} pure 1D models calibrated against the 3D simulations as
detailed in Section \ref{sect:patching} (un-patched models), and

\noindent{\bf PM:} those same 1D models, but with the surface layers
substituted by the averaged 3D simulations (patched models).

\noindent The two models are by construction identical interior to the 3D
simulations.
The frequency differences between PM and UPM will be the
asteroseismic surface effect due to convective effects on the average
stratification of the atmospheres.
This is in contrast to effects from the mode
dynamics through direct interactions between 3D convection and modes.
We refer to these two classes of seismic surface effects
as \emph{structure effects} and \emph{modal effects}, respectively.

The modal effects include the response to the pulsations of turbulent pressure and non-adiabatic energetics, including the convective flux.
\citet{houdek:3DsunSurfEff} computed modal components for the solar case,
based on a non-local, time-dependent mixing-length formulation of convection
\citep{gough:MLT-PulsStars}, and found them to be of the opposite sign 
and about 30\% of the
structure effects, bringing the total into remarkable agreement with
observations.
Using just the structural effects, as presented here, will therefore
give frequency shifts that are larger than the total seismic surface effect
(assuming that modal effects are always positive).
Computing modal components directly from the 3D simulations is a significant
project and will be the
subject of future papers.
The structure effect itself has two components, as detailed below.

\subsection{Turbulent pressure contribution}
\label{sect:Pturb}

The horizontally and temporally averaged (denoted by $\langle\dots\rangle$) turbulent pressure
\be{eq:pturb}
    p_{\rm t} = \langle\varrho u_z^2\rangle\ ,\quad{\rm with}\quad
    p = \langle p_{\rm g}\rangle + p_{\rm t}\ ,
\ee
contributes about half of the total convective expansion, where $\varrho$ is the
density, $u_z$ is the vertical velocity, $p_{\rm g}$ is the gas pressure and
$p$ the total pressure.
This expansion,
$\Lambda_{\rm t}$, by $p_{\rm t}$
can be directly quantified by integrating hydrostatic equilibrium
over just that component of the pressure
\be{eq:hydeq}
    \dd{p}{z} = g\langle\varrho\rangle \quad\Leftrightarrow\quad
    \Lambda_{\rm t}\equiv \Delta z_{\rm t} =
            \int\frac{{\rm d}p_{\rm t}}{g\langle\varrho\rangle}\ ,
\ee
where $z$ is the depth in the atmosphere.
This $\Lambda_{\rm t}$ is exact in the
sense that $\varrho$, $T$ and $p_{\rm g}$ do not change with $p_{\rm t}$, only
the location where those values occur are shifted.
The turbulent contribution to the total pressure,
$p_{\rm t}/p$, peaks at between 4\% for the coolest dwarf in
our grid, and 30\% for our warmest giant (see Fig.~\ref{turb_expan}),
just below the top of the convection zone.

\mfigur{turb_expan}{6.45cm,width=8.4cm}
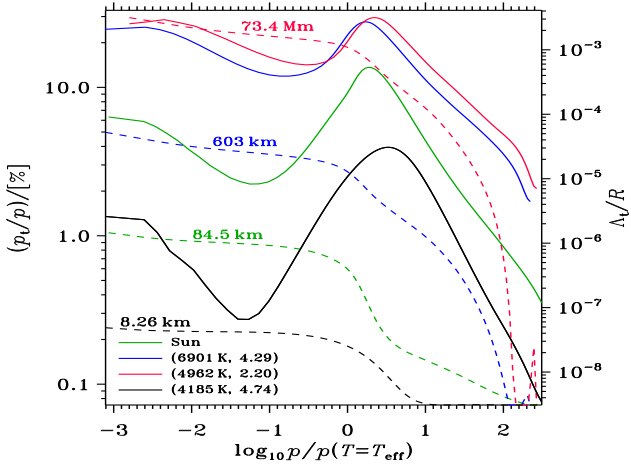
    {Ratios of turbulent to total pressures (solid lines, left-hand scale)
     for four simulations, spanning the range of behaviours in
     the grid of simulations, with $T_{\rm eff}$ and $\log g$ indicated in
     parenthesis. The right-hand axis shows the atmospheric
     expansion due to the turbulent pressure only, $\Lambda_{\rm t}$ (dashed lines), as fraction of stellar
     radius. The total $\Lambda_{\rm t}$ is indicated
     for each curve.}

The upturn in $p_{\rm t}/p$ above the photosphere, is not
convective but rather the effect of travelling waves escaping the acoustic
cavity above the acoustic cut-off frequency.
Notice that in local MLT formulations of convection, the convective
velocities would drop to zero from the peak of the $p_{\rm t}/p$-ratio in a
small fraction of a pressure scale-height, missing about half of the
atmospheric expansion from turbulent pressure.

\subsection{Convective backwarming}
\label{sect:ConvBackWarm}

Another, less straightforward, contribution to the convective expansion of
the atmosphere is caused by convective
fluctuations in the opacity. Since the top of convective envelopes occurs at
temperatures and densities where the opacity, $\kappa$, is extremely sensitive to
temperature (about $\kappa\propto T^9$ for the Sun) the convective temperature fluctuations
will cause much larger fluctuations in opacity. The warm upflows will be
shielded from cooling until (geometrically) close to the photosphere, as the
high opacity constitutes a geometric compression of the optical depth scale.
This effectively causes a warming below the photosphere, compared to a model
based on the opacity of the average stratification. The high power in $T$
means the opposing cooling effect in the downdrafts will be smaller.
The upflows also
occupy a larger fractional area. Coupled with the non-linear nature of
radiative transfer, the cooler downdrafts do not cancel the effect in the
upflows, resulting in a net warming below the photosphere. This in turn gives a
larger pressure scale-height and hence an expansion of the atmosphere, denoted
$\Lambda_\kappa$. The effect has a similar magnitude as that from the turbulent
pressure. The two effects are also correlated, as the amplitude of convective
velocities and temperature fluctuations are correlated. The total convective
expansion by the two mechanisms is denoted $\Lambda$ and is
shown in Fig.~5 of \citet{trampedach:3Datmgrid}. We compute this as the radial off-set of
pressure stratifications between the PM and UPM models, high in the atmosphere.

\subsection{Excluding the modal response of $p_{\rm t}$}

\citet{rosenthal:conv-osc} considered the effect of $p_{\rm t}$ on modes
using two simple cases as illustrative examples:

\noindent{\bf a): }
            $p_{\rm t}$ reacts exactly as $p_{\rm g}$, i.e., is in
            phase with the density fluctuations and proportional to them by
            $\gamma_1$.\\
\noindent{\bf b): }
            $p_{\rm t}$ has a completely incoherent response to modal
            density fluctuations, and over time 
            has no net effect on mode frequencies or eigen-functions, i.e.,
            exhibits no modal response.

\noindent Case b) result in Lagrangian pressure fluctuations
$\delta\ln p=(0\cdot p_{\rm t}/p+\gamma_1 p_{\rm g}/p)\,\delta\ln\varrho$,
where $\gamma_1$ is the adiabatic exponent of the gas, and
the parenthesis is referred to as the
\emph{reduced $\gamma_1$}. This should not be viewed as a reduction of the
thermodynamic quantity, but rather a statement about the turbulent pressure
response to modes.
Previous calculations \citep[e.g.,][]{piau:3D-SunSeism,sonoi:Sun3DSurfEffect,ball:SeismSurfExpanEff,magic:SolarSurfEff}
have all used case a), which is both an unjustified 
choice of the modal response to $p_{\rm t}$, as well as an incomplete
accounting of modal components.

\mfigur{freqdiff_fitBG14_Q20_t65g44m+00rf3D-t65g44m+00rf}{6cm}
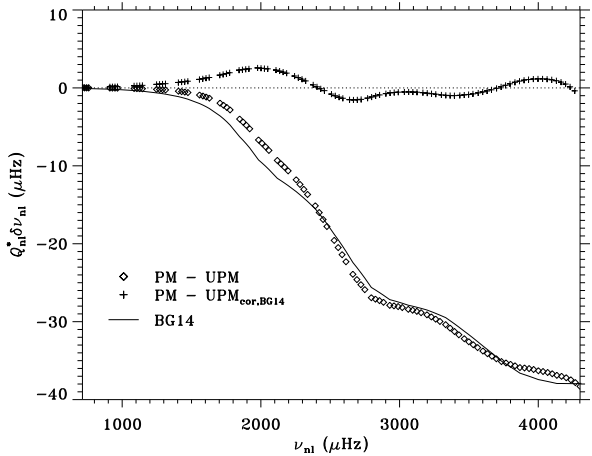
    {Scaled frequency differences in the sense: patched minus un-patched models
     (Sect.~\ref{sect:1D-3DSurfEff}) shown with $\diamond$,
     for $T_{\rm eff}=6$\,569\,K and $\log g=4.45$.
     We show both the two-term, BG14-fit (solid line), and the residual of that
     fit ($+$).}
To isolate the structural surface effect, we shall here use case b),
as did \citet{houdek:3DsunSurfEff},
to assume that there is no modal response to $p_{\rm t}$,
This results in a structural part of the surface effect which, for the solar
case, is about 1.4 times larger than the total surface effect at the acoustic cut-off frequency,
and about 3 times larger at $\nu_{\max}$ \citep[see][]{houdek:3DsunSurfEff}.

\section{Frequency shifts and discussion}
\label{sect:shifts}

We analyse the differences between adiabatic frequencies \citep[computed with ADIPLS;][]{jcd:ADIPLS} for the patched and
unpatched models, which we identify as the surface effect due to differences
in the average atmospheric structure of the two cases.
We do this for modes with degree $l=20$--$23$
and all orders, $n$, that have frequencies, $\nu_{nl}$,
below the acoustic cut-off frequency, $\nu_{\rm ac}$. This $l$-range ensures the modes
are confined well within the envelope models.

As first suggested  by \citet[][BG14]{ball:SurfEffFit}, we fit the frequency
differences, $\delta\nu_{nl}$, to expressions of the form
\be{eq:dnufit}
    I_{nl} \delta\nu_{nl} = c_{-1}\left(\nu/\nu_{\rm ac}\right)^{-1}
                          + c_3   \left(\nu/\nu_{\rm ac}\right)^3\ ,
\ee
where we evaluate the acoustic cut-off frequency as
\be{eq:numax}
    \nu_{\rm ac} = 5\,100\,\mu{\rm Hz\,}(g/g_\odot)
        \sqrt{T_{{\rm eff}\odot}/T_{\rm eff}}\ ,
\ee
scaled by the solar value \citep{jimenez:SolarNuCutoff}.
Equation (\ref{eq:dnufit})
was\linebreak[4] motivated by \citet{gough:HelioseismInfer}, exploring the origins
of the so-lar-cycle modulation of frequencies. He found that
a change of scale-height in the superadiabatic
layer would give rise to the first term,
while the $\nu^3$-term arises from a change to the
sound speed that keeps the density unchanged. In terms of the convective
expansion, these would arise from the convective backwarming and the turbulent
pressure, respectively.

The frequency shifts in Eq.~(\ref{eq:dnufit}) are scaled by the mode
inertia, $I_{nl}$
\citep[e.g.,][]{Asteroseismology}. This scaling renders the frequency shifts
independent of $l$, and likewise for the fit (to within 0.25\%).
This confirms that the restrictions on $l$, from using 
envelope models, do not limit the validity of our results.
Rather, it is an improvement, since a particular
atmosphere simulation can correspond to several interior models, in different
stages of evolution, potentially affecting the mode inertia. Our procedure
effectively separates the surface part, $c_{-1}$ and $c_3$, from the interior
part, $I_{nl}$ (supplied by the user), of $\delta\nu_{nl}$.
\mfigur{bgfit_terms_iscal_t65g44m+00rf3D-t65g44m+00rf}{6cm}
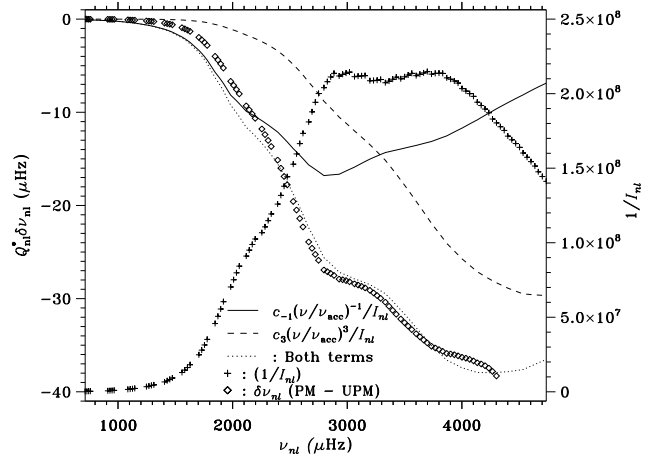
    {The various contributions to the fit to Eq.~(\ref{eq:dnufit}), for the
     case shown in Figure
     \ref{freqdiff_fitBG14_Q20_t65g44m+00rf3D-t65g44m+00rf}.
     The $\nu^{-1}$-term (solid) and the $\nu^3$-term (dashed), and the total
     fit (dotted) is compared to the actual frequency shift ($\diamond$).
     The inverse mode inertia ($+$) is also shown.}

In Figures \ref{freqdiff_fitBG14_Q20_t65g44m+00rf3D-t65g44m+00rf}--\ref{cwlgdnumaxnrm}, the frequency shifts
are reduced to $l=20$ by scaling with $Q^*_{nl}=I_{nl}/I_{20}(\nu_{nl})$, where
$I_{20}$ is $I_{n20}$ interpolated to $\nu_{nl}$.
\mfigur{cwlgdnumax}{8.2cm}
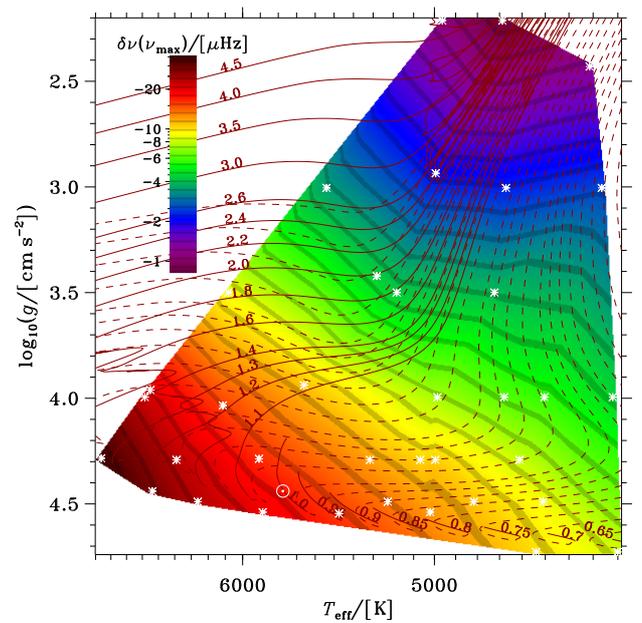
    {The amplitude of the frequency shift (patched minus unpatched) at $\nu_{\max}$, as
     function of $T_{\rm eff}$
     and $\log g$. The colour-scale is logarithmic.
     The location of the solar ($\odot$) and stellar ($*$) simulations are
     indicated in white.
     MESA \citep{paxton:MESA} stellar evolution tracks are over-plotted
     for masses as indicated.
     The dashed part shows the pre-main-sequence contraction.}
An example of our fit to Eq.~(\ref{eq:dnufit}) is shown in
Fig.~\ref{freqdiff_fitBG14_Q20_t65g44m+00rf3D-t65g44m+00rf} for a warm dwarf.
A power-law fit \citep{hans:SeismNearSurf} is obviously unable to fit the
frequency differences in Fig.~\ref{freqdiff_fitBG14_Q20_t65g44m+00rf3D-t65g44m+00rf}
over the full frequency range, as discussed by \citet{sonoi:Sun3DSurfEffect}.

How the two terms and the mode inertia contribute to the BG14-fit, is shown in
Figure \ref{bgfit_terms_iscal_t65g44m+00rf3D-t65g44m+00rf}. It is apparent
that the various bumps in the frequency shift are due to the mode inertia,
and the main reason the BG14 fit is so successful.

The amplitude of the surface effect at the frequency of maximum power,
estimated as $\nu_{\max} \simeq 0.6 \nu_{\rm ac}$, is shown in Figure
\ref{cwlgdnumax}.
This qualitatively agrees with analysis of {\it Kepler} observations
\citep{metcalfe:AMP42Kepler} of 42 F--G dwarfs and sub-giants. We performed
linear regression of their surface effects at $\nu_{\max}$, and of ours
interpolated to their targets, giving similar
increases with $\log T_{\rm eff}$ and $\log g$.
Our amplitudes are 2--8 times larger, however, partly due to our omission of
modal effects, 
expected to result in a surface effect larger than the total. Another important factor is
how stellar fits to seismic observations often exhibit coupled
parameters. In particular the surface effect, mixing length and helium content
can be strongly correlated, stressing the importance of constraining
these quantities independently.

Fig.~\ref{cwlgdnumax} shows that the magnitude of the surface effect increases
roughly as $g$. To take out this variation and highlight the relative
importance of the effect
we show in Figure \ref{cwlgdnumaxnrm} the fractional surface effect, in
units of $\nu_{\max}$.
This is seen to be predominantly, but not
exclusively, a function of the atmospheric expansion, based on the near
proportionality between Figure \ref{cwlgdnumaxnrm} and Figure 5 of
\citet{trampedach:3Datmgrid}.
\mfigur{cwlgdnumaxnrm}{8.2cm}
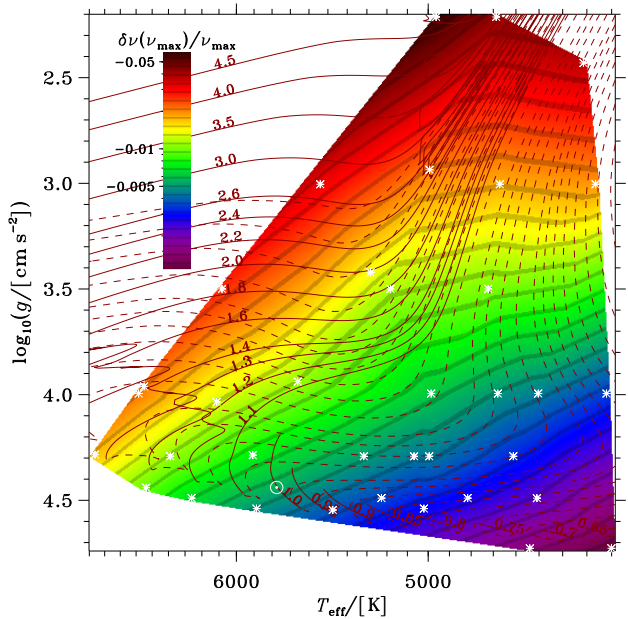
    {As Fig.~\ref{cwlgdnumax}, except the amplitude is
     normalised by $\nu_{\max}$,
     to highlight the relative significance of the
     surface effect.}
The ratio of the two terms in Eq.~(\ref{eq:dnufit}) at $\nu_{\max}$,
is shown in Figure \ref{cwcm1c3}, and illustrates a general change of shape
with atmospheric parameters. The $c_{-1}$-term dominates along a ridge
running parallel with the warm edge of our grid.
\mfigur{cwcm1c3}{8.2cm}
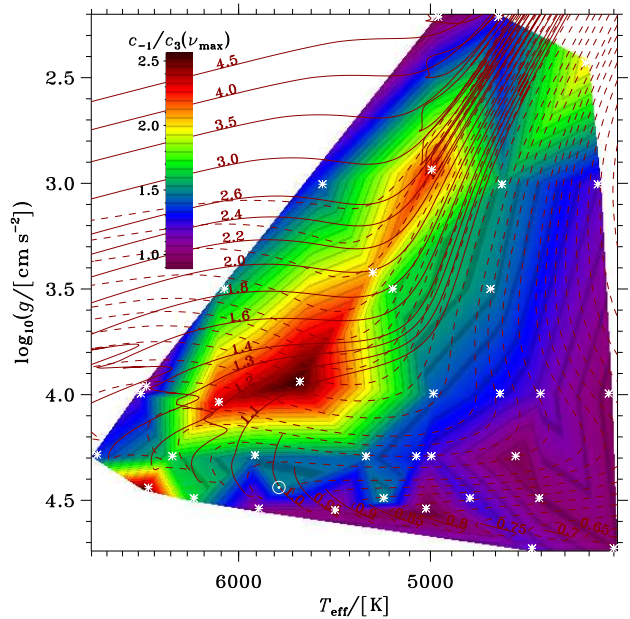
    {As Fig.~\ref{cwlgdnumax}, but showing the ratio of the two terms of
    Eq.\ [\ref{eq:dnufit}]) at $\nu_{\max}$, $(c_{-1}/c_3)\cdot(\nu_{\rm max}/\nu_{\rm ac})^{-4}$.}

We have evaluated the stuctural part of the asteroseismic surface
effect, as the effect on frequencies of the atmospheric expansion by realistic
3D convection, relative to 1D MLT stellar models.
Contrary to recent studies, we isolate the structural part from the modal part
of the surface effect by ignoring the turbulent pressure response
to modes, through the use of the so-called reduced $\gamma_1$. For the solar
case, this gives a frequency shift that is larger than the total, as the modal
part turns out to have the opposite sign.
Our results are well fit by BG14's expression, which also eliminates first-order
dependencies on $l$, so we can benefit from
using envelope models instead of full evolution models.

\section*{Acknowledgements}
We thank the referee for helpful comments.
RT acknowledges funding from NASA grant NNX15AB24G.
Funding for the Stellar Astrophysics Centre is provided by The Danish National 
Research Foundation (Grant DNRF106).


\begin{thebibliography}{23}
\expandafter\ifx\csname natexlab\endcsname\relax\def\natexlab#1{#1}\fi

\bibitem[{Aerts, Christensen-Dalsgaard \& Kurtz(2010)Aerts,
  Christensen-Dalsgaard, \& Kurtz}]{Asteroseismology}
Aerts C., Christensen-Dalsgaard J., Kurtz D.~W., 2010, Asteroseismology, 1st
  edn., Astron. \& Astroph. Library. Springer, Dordrecht

\bibitem[{Ball {et~al}\mbox{.}(2016)Ball, Beeck, Cameron, \&
  Gizon}]{ball:SeismSurfExpanEff}
Ball W.~H., Beeck B., Cameron R.~H., Gizon L., 2016, A\&A,
  \href{http://cdsads.u-strasbg.fr/abs/2016arXiv160602713B}{592, A159}

\bibitem[{Ball \& Gizon(2014)}]{ball:SurfEffFit}
Ball W.~H., Gizon L., 2014, A\&A,
  \href{http://cdsads.u-strasbg.fr/abs/2014A%26A...568A.123B}{568, A123}

\bibitem[{Brown(1984)}]{brown:SurfEff}
Brown T.~M., 1984, Science,
  \href{http://adsabs.harvard.edu/abs/1984Sci...226..687B}{226, 687}

\bibitem[{Christensen-Dalsgaard(2008{\natexlab{a}})}]{jcd:ASTEC}
Christensen-Dalsgaard J., 2008{\natexlab{a}}, Ap\&SS,
  \href{http://adsabs.harvard.edu/abs/2008Ap%26SS.316...13C}{316, 13}

\bibitem[{Christensen-Dalsgaard(2008{\natexlab{b}})}]{jcd:ADIPLS}
Christensen-Dalsgaard J., 2008{\natexlab{b}}, Ap\&SS,
  \href{http://adsabs.harvard.edu/abs/2008Ap%26SS.316..113C}{316, 113}

\bibitem[{Christensen-Dalsgaard, {D{\"a}ppen} \&
  Lebreton(1988)Christensen-Dalsgaard, {D{\"a}ppen}, \&
  Lebreton}]{jcd-Nature:osc-eos}
Christensen-Dalsgaard J., {D{\"a}ppen} W., Lebreton Y., 1988, Nature,
  \href{http://adsabs.harvard.edu/abs/1988Natur.336..634C}{336, 634}

\bibitem[{Christensen-Dalsgaard \& Frandsen(1983)}]{jcd-srf:stel-osc}
Christensen-Dalsgaard J., Frandsen S., 1983, Sol. Phys.,
  \href{http://adsabs.harvard.edu/abs/1983SoPh...82..469C}{82, 469}

\bibitem[{Gough(1977)}]{gough:MLT-PulsStars}
Gough D.~O., 1977, ApJ,
  \href{http://adsabs.harvard.edu/abs/1977ApJ...214..196G}{214, 196}

\bibitem[{Gough(1990)}]{gough:HelioseismInfer}
Gough D.~O., 1990, in Lecture Notes in Physics, Vol. 367, Progress of
  Seismology of the Sun and Stars, Osaki Y., Shibahashi H., eds., Springer,
  Berlin\href{http://cdsads.u-strasbg.fr/abs/1990LNP...367..283G}{, 283}

\bibitem[{Houdek {et~al}\mbox{.}(2017)Houdek, Trampedach, Aarslev, \&
  Christensen-Dalsgaard}]{houdek:3DsunSurfEff}
Houdek G., Trampedach R., Aarslev M.~J., Christensen-Dalsgaard J., 2017, MNRAS,
  \href{http://adsabs.harvard.edu/abs/2016arXiv160906129H}{464, L124}

\bibitem[{Hummer \& Mihalas(1988)}]{mhd1}
Hummer D.~G., Mihalas D., 1988, ApJ,
  \href{http://cdsads.u-strasbg.fr/abs/1988ApJ...331..794H}{331, 794}

\bibitem[{{Jim{\'e}nez}(2006)}]{jimenez:SolarNuCutoff}
{Jim{\'e}nez} A., 2006, ApJ,
  \href{http://adsabs.harvard.edu/abs/2006ApJ...646.1398J}{646, 1398}

\bibitem[{Kjeldsen, Bedding \& Christensen-Dalsgaard(2008)Kjeldsen, Bedding, \&
  Christensen-Dalsgaard}]{hans:SeismNearSurf}
Kjeldsen H., Bedding T.~R., Christensen-Dalsgaard J., 2008, ApJ,
  \href{http://adsabs.harvard.edu/abs/2008ApJ...683L.175K}{683, L175}

\bibitem[{Magic \& Weiss(2016)}]{magic:SolarSurfEff}
Magic Z., Weiss A., 2016, A\&A,
  \href{http://adsabs.harvard.edu/abs/2016A%26A...592A..24M}{592, A24}

\bibitem[{Metcalfe {et~al}\mbox{.}(2014)Metcalfe, Creevey, Do{\v g}an, Mathur,
  Xu, Bedding, Chaplin, Christensen-Dalsgaard, Karoff, \&
  Trampedach}]{metcalfe:AMP42Kepler}
Metcalfe T.~S. {et~al.}, 2014, ApJS,
  \href{http://adsabs.harvard.edu/abs/2014ApJS..214...27M}{214, 27}

\bibitem[{Paxton {et~al}\mbox{.}(2011)Paxton, Bildsten, Dotter, Herwig,
  Lesaffre, \& Timmes}]{paxton:MESA}
Paxton B., Bildsten L., Dotter A., Herwig F., Lesaffre P., Timmes F., 2011,
  ApJS, \href{http://adsabs.harvard.edu/abs/2011ApJS..192....3P}{192, 3}

\bibitem[{Piau {et~al}\mbox{.}(2014)Piau, Collet, Stein, Trampedach, Morel, \&
  {Turck-Chi{\`e}ze}}]{piau:3D-SunSeism}
Piau L., Collet R., Stein R.~F., Trampedach R., Morel P., {Turck-Chi{\`e}ze}
  S., 2014, MNRAS,
  \href{http://adsabs.harvard.edu/abs/2014MNRAS.437..164P}{437, 164}

\bibitem[{Rosenthal {et~al}\mbox{.}(1999)Rosenthal, Christensen-Dalsgaard,
  Nordlund, Stein, \& Trampedach}]{rosenthal:conv-osc}
Rosenthal C.~S., Christensen-Dalsgaard J., Nordlund {\AA}., Stein R.~F.,
  Trampedach R., 1999, A\&A,
  \href{http://adsabs.harvard.edu/abs/1999A%26A...351..689R}{351, 689}

\bibitem[{{Sonoi} {et~al}\mbox{.}(2015){Sonoi}, Samadi, Belkacem, Ludwig,
  Caffau, \& Mosser}]{sonoi:Sun3DSurfEffect}
{Sonoi} T., Samadi R., Belkacem K., Ludwig H.-G., Caffau E., Mosser B., 2015,
  A\&A, \href{http://adsabs.harvard.edu/abs/2015A%26A...583A.112S}{583}

\bibitem[{Trampedach {et~al}\mbox{.}(2013)Trampedach, Asplund, Collet,
  Nordlund, \& Stein}]{trampedach:3Datmgrid}
Trampedach R., Asplund M., Collet R., Nordlund {\AA}., Stein R.~F., 2013, ApJ,
  \href{http://cdsads.u-strasbg.fr/abs/2013ApJ...769...18T}{769, 18}

\bibitem[{Trampedach {et~al}\mbox{.}(2014{\natexlab{a}})Trampedach,
  Christensen-Dalsgaard, Nordlund, Asplund, \& Stein}]{trampedach:T-tau}
Trampedach R., Christensen-Dalsgaard J., Nordlund {\AA}., Asplund M., Stein
  R.~F., 2014{\natexlab{a}}, MNRAS,
  \href{http://cdsads.u-strasbg.fr/abs/2014MNRAS.442..805T}{442, 805}

\bibitem[{Trampedach {et~al}\mbox{.}(2014{\natexlab{b}})Trampedach,
  Christensen-Dalsgaard, Nordlund, Asplund, \& Stein}]{trampedach:alfa-fit}
Trampedach R., Christensen-Dalsgaard J., Nordlund {\AA}., Asplund M., Stein
  R.~F., 2014{\natexlab{b}}, MNRAS,
  \href{http://adsabs.harvard.edu/abs/2014MNRAS.445.4366T}{445, 4366}

\end{thebibliography}

\end{document}